\documentclass[12pt]{article} 
\usepackage{latexsym}
\usepackage{citesort,fullpage,amsfonts}
\begin{document}

\newsymbol \lesssim 132E
\newcommand{\di}{\partial}
\newcommand{\be}{\begin{equation}}
\newcommand{\ee}{\end{equation}}
\newcommand{\gh}{\hat{g}}
\newcommand{\sh}{\hat{s}} 
\newcommand{\al}{\alpha}
\newcommand{\bet}{\beta}
\newcommand{\Sig}{\Sigma}
\newcommand{\str}[1]{\frac{d#1}{d\hat{s}}}
\newcommand{\del}{\nabla}
\newcommand{\Gamh}{\hat{\Gamma}}
\newcommand{\xup}[2][\pm]{\ensuremath{\chi^{#2}_{#1}}}
\newcommand{\xdown}[2][\pm]{\ensuremath{\chi_{#1 #2}}}
\newcommand{\vup}[2][i]{\ensuremath{v^{#2}_{(#1)}}}
\newcommand{\vdown}[2][i]{\ensuremath{v_{(#1)#2}}}
\newcommand{\Seg}{\mbox{Segr\'{e} }}
\newcommand{\half}{\frac{1}{2}}
\newcommand{\third}{\frac{1}{3}}
\newcommand{\mR}{\ensuremath{\mathbb{R}}}

\newcounter{saveeqn}
\newcommand{\beqn}{\setcounter{saveeqn}{\value{equation}}%
	\stepcounter{saveeqn}\setcounter{equation}{0}%
	\renewcommand{\theequation}
	{\mbox{\arabic{section}.\arabic{saveeqn}\alph{equation}}}%
	\begin{eqnarray} }%

\newcommand{\eeqn}{\end{eqnarray}\setcounter{equation}{\value{saveeqn}}%
	\renewcommand{\theequation}{\arabic{section}.\arabic{equation}}}%

\newcommand{\mainlabel}[1]{\renewcommand{\theequation}%
	{\arabic{section}.\arabic{saveeqn}}\label{#1}%
	\renewcommand{\theequation}{\mbox{\arabic{section}.\arabic{saveeqn}%
	\alph{equation}}}}%

\title{Induced Matter and Particle Motion in Non-Compact Kaluza-Klein Gravity}
\author{ A. P. Billyard\thanks{Email: jaf@astro.queensu.ca} \\
		Department of Physics, \\
		Queen's University,\\
            	Kingston, Ontario, \\
		Canada K7L 3N6 \and 
		W. N. Sajko\thanks{Email: william.sajko@ca.jdsuniphase.com} \\
		Department of Physics, \\
		University of Waterloo, \\
		Waterloo, Ontario, \\
		Canada N2L 3G1 
	  } 
\maketitle
\baselineskip.4in
\begin{abstract}
\baselineskip.3in
We examine generalizations of the five--dimensional canonical metric
by including a dependence of the extra coordinate in the
four--dimensional metric.  We discuss a more appropriate way to
interpret the four--dimensional energy--momentum tensor induced from
the five--dimensional space-time and show it can lead to
quite different physical situations depending on the interpretation
chosen. Furthermore, we show that the assumption of five--dimensional
null trajectories in Kaluza--Klein gravity can correspond to either
four--dimensional massive or null trajectories when the path
parameterization is chosen properly.  Retaining the extra--coordinate
dependence in the metric, we show the possibility of a cosmological
variation in the rest masses of particles and a consequent departure
from four--dimensional geodesic motion by a geometric force.  In the
examples given, we show that at late times it is possible for
particles traveling along 5D null geodesics to be in a frame
consistent with the induced matter scenario.
\end{abstract}
\newpage
\renewcommand{\theequation}{\arabic{section}.\arabic{equation}}
\section{Introduction}
The modern version of non-compactified five--dimensional (5D)
Kaluza-Klein gravity, in which the 5D cylinder condition ($\di_4
\hat{g}_{AB}=0$)\footnote{Throughout this paper we use accent {\em
circumflex} to designate 5D quantities and no accents for 4D
quantities; also, uppercase Latin letters are used for the 5D
manifold, and lowercase Greek indices are used for the 4D manifold.
This paper uses units $8\pi G=c=1$ unless explicitly stated.}  has
been eliminated in favour of retaining the metric's dependence on the
extra coordinate, has had great success in describing
four--dimensional (4D) general relativity with an induced
energy-momentum tensor (see \cite{Overduin1997a} for a recent review).
The 5D space-time can be viewed as a foliation of 4D sheets on which
general relativity holds and a stress-energy tensor is induced through
the metric dependence on the extra coordinate \cite{Sajko1998a}.  This
procedure is always mathematically possible due to local embedding
theorems which state that a 4D Riemannian manifold (GR) can be locally
embedded in a 5D Ricci-flat Riemannian manifold
\cite{Rippl1995a,Romero1996a}.

In the induced--matter scenario, the induced Einstein tensor is
typically constructed from 4D metric $g_{\alpha\beta}$ defined by
\begin{equation}
d\hat{s}^2 = g_{\alpha\beta}(x^\Sigma,\ell)dx^\alpha dx^\beta +
\epsilon \phi(x^\Sigma,\ell)d\ell^2. \label{101}
\end{equation}
where the signature of the 4D metric $g_{\al\bet}$ is $(+,-,-,-)$;
also $x^\Sigma\equiv\{x^\alpha\}$, and $\epsilon \equiv \pm 1$, which
leaves the signature of the fifth dimension general and may allow a
``two-time'' metric (these types of metrics may appear odd but can be
shown to give sensible results in the induced-matter context
\cite{Billyard1996c,Billyard1996b}).  However, it has been shown
\cite{Mashhoon1994a} that metrics of the ``canonical form''
\begin{displaymath}
d\hat{s}^2 = \frac{l^2}{L^2}g_{\alpha\beta}(x^\Sigma)dx^\alpha
dx^\beta - d\ell^2
\end{displaymath}
lead to an induced false vacuum equation of state and hence this form
naturally leads to an induced cosmological constant, which is
parameterized by $L$.  Hence, it would seem that for manifolds of the form
\be
d\hat{s}^2 = \frac{l^2}{L^2}g_{\alpha\beta}(x^\Sigma,\ell)dx^\alpha
dx^\beta +\epsilon \phi(x^\Sigma,\ell)^2 d\ell^2,
\label{ouransatz}
\ee
part of the induced Einstein tensor would have a contribution from an
induced cosmological constant, an induced stress-energy from the
$\di_\ell g_{\alpha \beta}$ contributions as well as contributions
from the scalar field $\phi$.

Closely related to the induced-matter paradigm is the question of the
interpretation of 5D geodesics.  It has previously been shown
\cite{Billyard1997a} that if particles were to follow 5D geodesics,
then they cannot in general remain on $\ell=\ell_0$ hypersurfaces.
Therefore, the induced stress-energy tensor defined by $g_{\alpha
\beta}$ would not be what is observed by an observer moving along 5D
geodesics.  Within the Space-Time-Matter (STM) theory
\cite{Wesson1996a,Wesson1999a} to give a physically meaningful
interpretation to the extra coordinate, $\ell$ may be interpreted as
the rest mass of particles \cite{Wesson1984a} and so the change in the
rest--mass of a particle is dictated by the change in $\ell$.  Because
the induced matter is derived from a simple 5D theory, it is tempting
to assume that the motion of particles is also naturally 5D (in fact,
5D geodesic since the 5D manifold is a vacuum).  However, in general
this is incompatible with the induce--matter scenario.

In what follows we first derive the 4D induced energy--momentum tensor
from $g_{\alpha \beta}$ in (\ref{ouransatz}), decomposing it into a
false vacuum component, matter component and scalar field component
(if present).  In the literature, the induced matter is typically
interpreted as either a perfect fluid or a fluid with anisotropic
pressures, and we show that these are not the only possible types of
matter to model.  To demonstrate this we present two examples.  We
then explore the 5D null geodesic equation and show that these special
geodesics can reduce to 4D geodesics for massless particles, but there
is an acceleration of massive particles due to a geometric force
(which has been previously labeled as a ``fifth force''
\cite{Mashhoon1998a}) which depends on a scalar field and has an
explicit dependence on the extra dimension.  We then elucidate these
ideas with same two models and then make our final remarks.

\setcounter{equation}{0}

\section{4D Induced Matter From 5D Vacuum}
We wish to derive the induced matter resulting from the reduction of a
5D vacuum to a 4D hypersurface.  Consider the following gauge choice
for the 5D metric which explicitly depends on the extra coordinate
$x^4\equiv \ell$, and for which $\hat{g}_{\alpha 4}\propto A_\alpha$ (the
electromagnetic vector potential) is set to zero.  We factor out a
conformal dependence on the 4D metric and include a scalar field so
that the 5D metric can be written as
\be
\label{201}
\gh_{AB} =  \left( \begin{array}{cc}
                           \frac{\ell^2}{L^2}\,g_{\al\bet}(x^\Sig,\ell) & 0 \\
                           0      & \epsilon \phi^2(x^\Sig,\ell)
                           \end{array} \right) \, .
\ee 

The easiest way to  determine the induced matter on the 4D
hypersurfaces  ($\ell=\ell_o=const.$)  is to decompose the 5D metric using a 4+1  
decomposition;  the ``4'' is used to designate  4D
hypersurfaces with an induced metric $ (\ell_o^2/L^2) g_{\al\bet}$, and the ``1''
corresponds to the lapse in the extra
dimension between adjacent 4D hypersurfaces measured by the
scalar field $\phi$.  This procedure was initially
performed in \cite{Sajko1998a}, and for the metric
(\ref{101}) the
components of the 5D vacuum field equations $\hat{R}_{AB}=0$ are:
\beqn
\label{202}
\hat{R}_{\al\bet}=0 &\Rightarrow&  R_{\al\bet}=\frac{1}{\phi}
	\del_\al \del_{\bet} \phi - \frac{\epsilon}{\phi} \di_{\,\ell} 
	K_{\al\bet} +\epsilon \left(K K_{\al\bet}-2 K_{\al\gamma} 
	{K^{\gamma}}_{\bet}\right)\, , \label{203} \\
\hat{R}_{\ell\beta} = 0 &\Rightarrow & \del_\al \left({K^{\al}}_{\bet}- 
	{\delta^{\al}}_{\bet} K \right)=0 \, ,\label{204} \\
\hat{R}_{\ell\ell} =0 &\Rightarrow & \epsilon \,\Box \phi =\di_{\,\ell} K
                    - \phi K^{\al\bet} K_{\al\bet} \,,
\eeqn
where the covariant derivative and the d'Alembertian operator ($\Box$)
are defined on the 4D hypersurfaces.  Here the extrinsic curvature of
the embedded 4D hypersurfaces is defined as
\be
\label{205}
K_{\al\bet} \equiv -\frac{1}{2\phi} \di_{\,\ell} \left( \frac{\ell^2}{L^2} 
              g_{\al\bet} (x^\Sigma,\ell)  \right) \, ,
\ee       
and $K\equiv K^\alpha_\alpha=\frac{L^2}{\ell^2}g^{\alpha\beta}K_{\alpha\beta}$.
It is evident that the extra coordinate dependence in the 4D metric
plays a crucial r\^{o}le in inducing matter in 4D.  However, if 
 $\di_{\,\ell}\,g_{\al\bet}=0$ then the only consistent
solution to the above equations is
\be
\label{206}
\di_{\,\ell}\,g_{\al\bet} =0 \quad \Rightarrow \quad 
R_{\al\bet}=\frac{3\epsilon}{L^2}\, g_{\al\bet}\, ,\quad
\phi=1. 
\ee
This solution can be identified as a false vacuum (i.e., $\mu =-p = \Lambda$), 
provided the constant $L$ is identified with $\Lambda$ via
\be
\label{207}
\Lambda\equiv-\frac{3 \epsilon}{L^2} \, . 
\ee
The induced cosmological constant generates either the de Sitter
vacuum when $\epsilon=-1$ ($\Lambda > 0$) or the anti-de Sitter vacuum
when $\epsilon=+1$ ($\Lambda < 0$, which leads to a two-time metric).
When the 4D metric depends on $\ell$ the extra terms generated by the
derivatives with respect to the extra coordinate (and possibly the
scalar field terms) can be viewed as the matter contribution to the
stress-energy, whereas terms proportional to $g_{\al\bet}$ can be
related to the vacuum stress-energy.

Let us now investigate the matter induced from the energy--momentum
 tensor derived from $g_{\al\bet}(x^\Sigma, \ell)$, assuming that
 $\phi=\phi(x^\Sigma)$.  First, we isolate terms in (\ref{202})
 proportional to $g_{\al\bet}$ and identify these terms with the
 induced effective cosmological ``constant'',
 $\Lambda_{\mbox{{\scriptsize \em eff}}}$.  Therefore, we begin with
\be
K_{\al\bet}=-\frac{\ell}{\phi L^2} g_{\al\bet}-\frac{\ell^2}{L^2}J_{\al\bet},
\label{Kbreak}
\ee 
where 
\be
J_{\al\bet} \equiv \frac{1}{2\phi}\di_\ell g_{\al\bet}.
\ee
Substituting (\ref{Kbreak}) into (\ref{202}) leads to
\be
R_{\al\bet}=\frac{\del_\al \del_{\bet} \phi}{\phi}
	 +\frac{3\epsilon}{\phi^2 L^2}\left(1+\third\phi \ell J
	 \right)g_{\al\bet} + \frac{\epsilon \ell^2}{\phi L^2} \left(
	 \frac{4J_{\al\bet}}{\ell} + \di_\ell J_{\al\bet} 
	+ \phi(J J_{\al\bet} - 2 J_{\alpha\gamma} J^\gamma_\bet) \right), 
\ee
(where $J\equiv J^\alpha_\alpha=g^{\alpha\beta}J_{\alpha\beta}$) and
hence, the effective cosmological constant is defined as
\be
\label{29}
\Lambda_{\mbox{{\scriptsize \em eff}}} = - \frac{3\,\epsilon}{\phi^2
L^2}\left(1+\frac{\ell}{6}g^{\mu \nu}\di_\ell g_{\mu\nu} \right).  
\ee
The induced Einstein field equations can thus be 
written
\be G_{\al \bet} = \!\,^{(\phi)}T_{\al \bet}
		+\Lambda_{\mbox{{\scriptsize \em eff}}}~ g_{\al\bet} 
		+\frac{\!\,^{(M)}T_{\al \bet}}{\phi},
\label{induced}
\ee
where
\beqn
\mainlabel{breakdown}
\!\,^{(\phi)}T_{\al \bet} &=& \frac{\del_\al \del_{\bet} \phi}{\phi} -\frac{\Box\phi}{\phi}g_{\al\bet},\\
\nonumber
\!\,^{(M)}T_{\al \bet} &=& \frac{\epsilon \ell^2}{L^2} \left\{
	 \frac{4J_{\al\bet}}{\ell} +
	 \di_\ell J_{\al\bet} + \phi\left(J  J_{\al\bet} - 2 J_{\alpha\gamma} J^\gamma_\bet\right)\right. \\
&& \left. -\half g_{\al\bet}\left[ \frac{6J}{\ell} +
	 g^{\mu\nu}\di_\ell J_{\mu\nu}+\di_\ell J + \phi\left(J^2
	- J^{\mu\nu} J_{\mu\nu}\right)\right]\right\}.\label{The_Matter}
\eeqn
Note that five--dimensional vacuum relativity corresponds to a
$\omega=0$ Brans--Dicke theory \cite{Billyard1997a}, which is why we
have left an explicit factor of $\phi^{-1}$ in front of the matter
term in (\ref{induced}).  The case $\phi=1$ reduces to ordinary 4D relativity
with matter.

It is necessary to comment on the kinematic quantities of
$^{(M)}T^\al_\bet$.  Often in the literature concerning induced matter
from Kaluza--Klein theory, it is often assumed that the induced
stress--energy tensor represents either perfect fluid model or a fluid
model with anisotropic pressures.  However, this is not necessarily
the case; indeed the induced stress--energy tensor may not be
appropriate for a fluid source at all.  To represent a fluid source,
the tensor $^{(M)}T^\al_\bet$ must be of \Seg type \{1,1,1,1\}; that
is, in its Jordan form, $^{(M)}T^\al_\bet$ will be diagonal, the
components of which will be the eigenvalues of the energy--momentum
tensor.  One eigenvalue will be associated with a time--like
eigenvector and the other three will be associated with space--like
eigenvalues.  If this is satisfied, then $^{(M)}T^\al_\bet$ can be
modeled as a fluid with a time--like velocity field $u^\al$.  If the
space--like eigenvectors are all equal then, {\em and only then}, can
the stress tensor be modeled as a perfect fluid.  The kinematic
quantities $\{\mu, p, u^\al, q^\al, \pi^\al_\bet\}$\footnote{Here,
$\mu$ is the fluids energy density, $p$ its averaged pressure,
$q^\alpha$ is the fluids heat conduction vector and $\pi^\alpha_\beta$
is the fluids anisotropic pressure tensor.} can thus be determined
from the eigenvalues and eigenvectors, and Appendix \ref{eigen}
describes how to compute these quantities for two important cases:
fluids with heat conduction and isotropic pressures, $q^\al\neq 0$ \&
$\pi^\al_\beta=0$, and fluids without heat conduction  $q^\al= 0$
\& $\pi^\al_\beta\neq0$.  We now present two examples.

\subsection{Example A: Ponce de Leon metric}

The first example is the one-parameter class of
solutions found by Ponce de Leon \cite{Ponce1988a}:
\be
\label{404}
d\sh^2=\frac{\ell^2}{L^2}\left[ dt^2-
\left(\frac{t}{L}\right)^{2/\al}\,\left(\frac{\ell}{L}\right)^{2\al/(1-\al)}\,
d\vec{x} \cdot  d\vec{x}\right]
-\left(\frac{\al}{1-\al}\right)^2\,\left(\frac{t}{L}\right)^2\, d\ell^2 \, ,
\ee
where $\alpha$ is a constant.  These solutions have been
previously used in a cosmological context since they are spatially
isotropic and homogeneous, and  on the induced 4D hypersurfaces they are
the analogues of the $k=0$ FRW cosmologies.  Using (\ref{29}) for the
effective cosmological constant and (\ref{induced}) for the induced
stress-energy tensors we find that
\beqn
\Lambda_{\mbox{{\scriptsize \em eff}}} &= &\frac{3(1-\al)}{(\al t)^2}\\ \nonumber \\
\mbox{total stress-energy:}\qquad
T^\al_\bet &=& \left[\begin{array}{cc}\frac{3}{(\al t)^2} & 0 \\
				   0    & -\frac{(2\al-3)}{(\al t)^2}
				{\delta^i}_j\end{array}\right]\\ \nonumber
\\
\mbox{scalar stress-energy :}\qquad
\label{Tphi}
\!\,^{(\phi)}T^\al_\bet &=& \left[\begin{array}{cc}\frac{-3}{2\al t^2} & 
					0 \\
				   0    & -\frac{1}{2\al t^2}
				{\delta^i}_j\end{array}\right]\\ \nonumber 
\\
\mbox{matter stress-energy :}\qquad
\!\,^{(M)}T^\al_\bet &=&
	\left[\begin{array}{cc}\frac{9}{2(1-\al)Lt} & 	0 \\
	0    & \frac{3}{2(1-\al)Lt} {\delta^i}_j\end{array}\right],\label{Mse}
\eeqn
where ${\delta^i}_j$ is the three--dimensional Kronecker--delta
function.  We note that the effective cosmological constant decreases
as $t^{-2}$ which is compatible with string inspired cosmological
theories \cite{Lopez1996a} and scalar-tensor gravity \cite{Endo1977a} (for an
extensive bibliography on variable $\Lambda$ cosmologies see
\cite{Overduin1998a}).  This is favourable for inflationary models since
the cosmological term is large for early times and then decreases to
zero for late times.  As is evident from the induced energy--momentum
tensor (\ref{Mse}), all three space--like eigenvalues are equal and so
$\,\!^{(M)}T^\al_\bet$ can aptly represent a perfect fluid with the
energy--density ($\mu$) and pressure ($p$) given by
\be
\mu = \frac{9}{2 (1-\al)Lt}, \quad p= -\frac{3}{2 (1-\al)Lt}  \qquad 
\Longrightarrow \qquad p=-\third \mu . \label{justmatter}
\ee
Here we see the fluid behaves like a barotropic fluid with a linear
equation of state parameter $\gamma=2/3$.  Note that it is necessary
to impose that $\al\leq 1$ ($\mu\geq 0$) in which case the effective
cosmological constant is positive. Furthermore, one could demand
that the stress-energy tensor arising from the scalar field (\ref{Tphi}) satisfies
the energy conditions (weak, strong and dominant), in which case
$\alpha\leq 0$.


If the {\em entire} energy--momentum tensor is treated as one
fluid, we obtain
\be
\label{405}
\mu_{\rm tot} = \frac{3}{ (\al t)^2}, \quad 
p_{\rm tot}= \frac{(2\al-3)}{(\al t)^2} \qquad \Longrightarrow \qquad 
p_{\rm tot} = \left(\frac{2}{3}\al-1\right)\,\mu_{\rm tot} .\label{all}
\ee
which is consistent with that found in \cite{Wesson1992a}.  In this
case, we have a barotropic fluid with a linear equation of state
parameter $\gamma=\frac{2}{3}\al$.  The strong energy condition
($\mu_{\rm tot}+3p_{\rm tot}\geq0$) restricts $\alpha\geq 1$ while the dominant energy
condition ($\mu_{\rm tot}\geq |p_{\rm tot}|$) restricts $0\leq\alpha\leq 3$. As discussed
in \cite{Wesson1992a}, there are three physically relevant choices for
$\al$: $\al \in (0,1)$ for inflation, $\al = 2$ for radiation, and
$\al = 3/2$ for dust. For the latter two values, the cosmological
constant is negative ($\al= 0,1$ are bifurcation values and must be
treated separately).

We present this second interpretation (\ref{405}) to demonstrate how
different the induced matter can be when we consider the
stress--energy tensor as a conglomerate of three separate sources, but
feel the first interpretation (\ref{justmatter}) is more appropriate.
First of all, such a decomposition is consistent with how the
five--dimensional vacuum theory is mathematically equivalent to
four--dimensional Brans--Dicke theory (with or without a cosmological
constant).  Secondly, by considering the scalar field as a separate
source, problems such as the discrepancy between gravitational and
inertial mass can be resolved (see, for example, \cite{Sajko1999b}).

\subsection{Example B: Shell--like Solutions}
The next example is a two-parameter class of spherically symmetric
solutions \cite{Wesson1998a}: 
\be \label{Shell}
d\sh^2=\frac{\ell^2}{L^2}\left(A^2 dt^2-B^2dr^2-C^2r^2d\Omega^2\right)
- d\ell^2,
\ee
where
\beqn
\mainlabel{coefficients}
A&=& \frac{1}{B}+\frac{k_2L}{\ell}, \\
B&=& \frac{1}{\sqrt{1-\frac{r^2}{L^2}}}, \\
C&=& 1+\frac{k_3L^2}{r\ell},
\eeqn
(note that this form can be expressed in the original form given in
\cite{Wesson1998a} by letting $k_2\rightarrow k_2/k_1$ and
$t\rightarrow k_1 t$).  Since these models have $\phi=1$ they
correspond to 4D relativistic models (as opposed to 4D $\omega=0$
Brans--Dicke models).  These solutions have been termed ``shell''
solutions since at
\begin{equation}
r=r_C = \frac{|k_3|L^2}{\ell}
\end{equation}
(where $C(r_C)=0$) the density and pressure of the fluid found in
\cite{Wesson1998a} diverged (at $r=r_C$ the surface area of the two
sphere, $4\pi r^2 C^2$, is zero and so this may be taken as the origin
of the system), and at \begin{equation}
r=r_A = L\sqrt{1-\frac{k_2^2L^2}{\ell^2}},
\end{equation}
(where $A(r_A)=0$) the pressure diverges.  Note that $r_A$ and $r_C$ coincide at
\beqn
\mainlabel{horizon_limit}
r_0 &=&\frac{\left|k_3\right|}{\sqrt{k_2^2+k_3^2}}L \\
\ell_0 &=& \sqrt{k_2^2+k_3^2}~L,
\eeqn
and so for $\{r,\ell\}<\{r_0,\ell_0\}$ we have that $r_A<r_C$; since $r_C$ is defined as
the centre of the system, $r_A$ is excluded from the manifold for
$\ell<\ell_0$.

The four--dimensional component of this metric is the de Sitter metric when the
parameters $k_2$ and $k_3$ are both zero; thus because of the
dependence on the extra-coordinate, this metric may be interpreted as
a generalization to the de Sitter vacuum with an effective
cosmological constant.  To preserve the signature, the radial
coordinate must obey $r < |L|$.  Furthermore, we adopt the assumptions
used in \cite{Wesson1998a} that $L>0$, $\ell>0$, $k_2<0$ and $k_3<0$.

Now, equations (\ref{29})-(\ref{breakdown}) reduce to:
\beqn
G^\al_\bet &=& {\rm diag}\left[
	\frac{(1+2C)}{L^2C^2},\frac{(AB+2C)}{L^2ABC^2},
	\frac{C+AB+1}{L^2ABC},\frac{C+AB+1}{L^2ABC}\right], \\
\Lambda_{\mbox{{\scriptsize \em eff}}} &= &\frac{C+2AB}{L^2ABC}, \\
\!\,^{(M)}T^\al_\bet &=& \!\!{\rm diag}\left[
	\frac{(AB-C^2)}{L^2ABC^2},\frac{(AB-C^2)+2C(1-AB)}{L^2ABC^2},
	\frac{(1-AB)}{L^2ABC},\frac{(1-AB)}{L^2ABC}\right]. \label{Mse2}\nonumber \\ 
\eeqn
Clearly, the eigenvalues of the induced energy--momentum tensor
(\ref{Mse2}) are (see appendix)
\beqn
\lambda_+&=& \frac{(AB-C^2)}{L^2ABC^2},\\
\lambda_-&=& \frac{(AB-C^2)}{L^2ABC^2} +2\frac{(1-AB)}{L^2ABC},\\
\lambda_2=\lambda_3 & =& \frac{(1-AB)}{L^2ABC}.
\eeqn
At this point, one can model $\!\,^{(M)}T^\al_\bet$ as an imperfect
fluid, but there is no unique choice.  However, because
$\lambda_2=\lambda_3$ there are two obvious models from which to
choose: a fluid with heat conduction and isotropic pressures
($q^\alpha\neq 0$, ${\pi^\alpha}_\beta=0$) and a fluid with no heat
conduction and anisotropic pressures ($q^\alpha= 0$,
${\pi^\alpha}_\beta\neq 0$).

\subsubsection{Heat Conduction with Isotropic Pressure}
For the case $q^\alpha\neq 0$ and  ${\pi^\alpha}_\beta=0$, equations
(\ref{heateqns}) in 
the appendix lead to the following kinematic quantities:
\beqn
\mu & = & \left(\frac{1-AB}{L^2ABC}+2\frac{AB-C^2}{L^2ABC^2}\right), \\
p & = & -\left(\frac{1-AB}{L^2ABC}\right), \\
u^\alpha & = & \frac{1}{\sqrt{2(1-AB)C}}
	\left[-\frac{\sqrt{(C+1)(C-AB)}}{A}, 
	     \frac{\sqrt{(C-1)(C+AB)}}{B}, 0, 0\right], \\ \nonumber \\
q^\alpha & = &
\frac{\sqrt{(C^2-1)(C^2-A^2B^2)}}
	{L^2AB\sqrt{2(1-AB)C^5}}\left[-\frac{\sqrt{(C-1)(C+AB)}}{A}, 
	\frac{\sqrt{(C+1)(C-AB)}}{B}, 0, 0\right], \nonumber \\ \\
q^2 & = &\frac{(1-C^2)(A^2B^2-C^2)}{(L^2ABC^2)^2}
\eeqn
(where it can be verified that $u^\al u_\al=1$).  As discussed in the
appendix, $\!\,^{(M)}T^\al_\bet$ will be of \Seg type \{1,1,1,1\} if
$(\mu+p)^2-4q^2>0$ and indeed
\be
(\mu +p)^2-4q^2 = \frac{4(AB-1)^2}{(L^2ABC)^2}>0.
\ee

\subsubsection{Anisotropic Pressure with No Heat Conduction}
For the case where $q^\al=0$ and $\pi^\al _\beta\neq0$, equations
(\ref{noheateqns}) yield
\beqn
\mu  & = & \frac{(AB-C^2)}{L^2ABC^2},\\
p   & = & -\third\frac{AB-C^2}{L^2ABC^2}-\frac43\frac{(1-AB)}{L^2ABC},\\
u^\al & = & \left[\frac{1}{A},0,0,0\right] \label{anivelo}\\
\pi^\al_\beta & = &\left[ \begin{array}{cccc} 
	0 & 0 & 0 & 0 \\
	0 & -\frac23\frac{(C-1)(C+AB)}{L^2ABC^2}& 0& 0 \\
	0 & 0 & \third\frac{(C-1)(C+AB)}{L^2ABC^2} & 0 \\
	0 & 0 & 0 & \third\frac{(C-1)(C+AB)}{L^2ABC^2}
	\end{array}\right].
\eeqn

Clearly, as sections 2.2.1 and 2.2.2 demonstrate, the same 5D metric
can yield two very different physical models.  In the first case the
induced matter is that of a fluid which has heat conducting in the
radial direction, whereas the second case is an induced matter without 
heat conduction but with anisotropic pressures (the radial pressure is
different from the solid angle pressure).  However, these are not the
only possible models from which to choose.  For instance, in
\cite{Hall1984a} it has been shown that stress tensors of \Seg type
$\{1,1,(1,1)\}$ can also be used to model a perfect fluid and a
electromagnetic field (either null or non--null). 

\setcounter{equation}{0}

\section{Motion, Mass Variation, and the Geometric Force}

In this section we approach particle dynamics from a 5D Lagrangian for
the canonical metric (\ref{201}) and use the Euler-Lagrange equations
to obtain the acceleration equation induced in 4D.  When the path
parameterization is chosen judiciously we show that the components of
the 5D acceleration equation reproduce the 4D geodesic equation for
null particles and an acceleration equation for massive particles.
With the interpretation is that the extra coordinate is related to the
rest-masses of particles \cite{Wesson1984a} the 5D null
geodesics lead to a rest-mass variation for massive particles. We
elucidate these results with the models studied in 2.1 and 2.2.

\subsection{Motion and Mass Variation}
To study dynamics in 5D Kaluza-Klein gravity with the canonical metric
(\ref{201}) we begin by extremizing the action
\be
\label{c27}
 \hat{I} = \int\limits^B_A \hat{{\cal L}}(x^A, \dot{x}^A)\, d\lambda
 =\int\limits^B_A d\lambda ~\sqrt{\frac{\ell^2}{L^2} g_{\al\bet}(x^\Sig,\,\ell)
       \frac{dx^\al}{d\lambda}\frac{dx^\bet}{d\lambda}
      + \phi^2(x^\Sig,\,\ell) \frac{d\ell^2}{d\lambda^2}} ,
\ee
where $\lambda$ is an arbitrary path parameter and the velocities are
coterminal at the points $A,B$.  With these boundary conditions,
extremizing the action gives the well-known Euler-Lagrange equations
\be
\label{c28}
\frac{d}{d\lambda}\left(\frac{\di \hat{{\cal L}}}{\di \hat{u}^A}\right) -
\frac{\di \hat{{\cal L}}}{\di x^A} = 0 \quad \Rightarrow \quad
\frac{d \hat{u}^A}{d\lambda}+\Gamh^{A}_{BC} \,\hat{u}^B \,\hat{u}^C = \hat{u}^A
\frac{d}{d\lambda}\left(\ln \hat{{\cal L}}\right)\, .
\ee
The 4D and $\ell$ components of equation (\ref{c28}) are 
\beqn
\mainlabel{29parts}
\label{c29}
&& u^\beta \del_\bet u^\alpha =
\frac{d}{d\lambda}\ln\left(\frac{\hat{{\cal L}}}{\ell^2}\right) u^\alpha
- g^{\alpha\beta}\left[\di_{\,\ell}g_{\beta\gamma} u^\gamma
+\frac{1}{2}\left(\frac{L\phi}{\ell}\right)^2 \di_\beta \left( \ln
\phi^2 \right) \dot{\ell} \right] \dot{\ell} \\  \nonumber \\
\label{c30}
&& \left(\frac{L\phi\dot{\ell}}{\ell}\right) \left\{\Upsilon
\left[\ln\left(\frac{L\phi\dot{\ell}}{\ell}\right)\right]^{\dot{}}
+\frac{\dot\phi}{\phi}\left[\Upsilon-\left(\frac{L \phi \dot{\ell}}{\ell}\right)^2 \right]
\right\} \nonumber\\
&& \qquad = 
\frac{1}{2}\left[\Upsilon 
- \left(\frac{L
\phi\,\dot{\ell}}{\ell}\right)^2\right]\left[\frac{2\Upsilon}{\ell}   
+\di_{\,\ell} g_{\alpha\beta} u^\alpha u^\beta
- \left(\frac{L\phi\dot{\ell}}{\ell}\right)^2 \di_{\,\ell} \ln \phi^2
\right],\nonumber \\
\eeqn
where a dot is shorthand for $d$/$d \lambda$.  If the
parameterization, $\lambda$, were chosen to be either the 5D proper
distance, $\hat{s}$, or a 5D null parameterization, then term on the
right hand side of (\ref{c28}) vanishes (and hence (\ref{c28})
describe 5D geodesics); however, we have chosen the parameterization
to be the 4D proper distance, $\lambda=s$, so that $u_\al
u^\al\equiv\Upsilon$ (where $\Upsilon=1$ for timelike paths and
$\Upsilon=0$ for null paths).  The extra terms on the right hand side
of the equations (\ref{29parts}) are a consequence of this choice
rather than the 5D proper distance $\lambda=\sh$ \cite{Mashhoon1998a}.
Solving equation (\ref{c30}) for $\dot{\ell}$ is very complicated, and
in general the quantities $\{g_{\alpha \beta},\phi\}$ would have to be
first specified.  However, from (\ref{c30}) it is apparent that the
solution
\be
\label{c31}
\left(\frac{\dot{\ell}}{\ell}\right)^2=\frac{\Upsilon }{L^2 \phi^2}\, 
\ee
satisfies (\ref{c30}) identically for any $\{g_{\alpha \beta},\phi\}$.
It may be verified that (\ref{c31}) represents 5D null geodesics by
examining the 5D canonical line element (\ref{ouransatz}).  Hence, the
particle paths are consequently 5D null even though we have chosen the
4D proper distance $\lambda=s$ to be the path parameter.  

Relation (\ref{c31}) constrains the velocity $\dot{\ell}$ but does not
give it physical meaning; for this, we turn to Kaluza-Klein theories
in which the extra coordinate can be interpreted as a geometric mass
via $\ell=G m/c^2$
\cite{Overduin1997a,Wesson1984a,Wesson1996a,Wesson1997a,Wesson1999a}.
We now look at the variation of rest mass as a function of the 4D path
parameterization.  The rest mass of a particle is easily obtained from
integrating (\ref{c31}):
\be
\label{c32}
m=m_o \exp\left(\pm \sqrt{\frac{\Upsilon}{L^2}} \int ds \,
\phi^{-1}\right).
\ee
Since in 4D we have $\Upsilon=0$ for photons, this implies that the
variation in a photon's rest mass is zero and so its mass
may consistently be set to zero.  However, for 4D paths which have
$\Upsilon=1$,  there is  a variation in the rest-mass of massive particles
driven by the
scalar field $\phi$ and hence $\phi$ may be modeled as a Higgs-type field. 
Let us make a few comments:
\begin{enumerate}
\item A conformal transformation of the 4D metric $g \rightarrow
\tilde{g} = \phi^2 g$ would remove the scalar-field dependence in
(\ref{c32}), but also changes the induced-matter field equations as
well as the 4D acceleration $a^\al=u^\mu\del_\mu
u^\alpha$, and complicates matters substantially.
\item When the 4D condition $\di_{\,\ell}\, g_{\al\bet}=0 \rightarrow
\phi=1$ is imposed, we get a cosmological variation of the rest masses
of massive particles in the de Sitter vacuum ($\epsilon=-1$), namely that
\be
m=m_0e^{\pm (s-s_0)/L}.
\ee
\item If we choose a two-time metric ($\epsilon=+1$) the
variation is imaginary, giving an oscillating rest mass in the anti-de
Sitter vacuum (this oscillation will hold even for more complicated
metrics which do not obey $\di_{\,\ell}\, g_{\al\bet}=0$).

\end{enumerate}
We now turn our attention to the acceleration equation (\ref{c29}).
After some algebra, equation (\ref{c29}) reduces to the form
\be
\label{c33}
u^\bet \del_\bet\, u^\al = f^\al\, ,
\ee
where $f^\al$ is  the force per unit rest mass
\be
\label{c34}
f^\al=- h^{\al\gamma}\left(\Upsilon  \frac{ \phi_\gamma}{\phi}+
          \di_{\,\ell}\, g_{\gamma\bet} \,u^\bet\, \dot{\ell}\right) \,,
\ee
and $ h^{\al\gamma}\equiv g^{\al\gamma}-u^\al u^\gamma$ is the
projection tensor.  When $\di_{\,\ell}\, g_{\al\bet}=0$ ($\phi=1$),
this force term vanishes, and the motion is geodesic for both photons
and massive particles in a pure 4D de Sitter vacuum, which is the
correct 4D result in general relativity.  However, when
$\di_{\,\ell}\, g_{\al\bet}\neq0$, photons will still travel along
null 4D geodesics since they obey $\Upsilon=0$ and $\dot{\ell}=0$; but
massive particles will experience a geometric force since $\Upsilon=1$
and $\dot{\ell}\ne 0$. We now consider some examples to elucidate
these ideas.
   
\subsection{Example A: Ponce de Leon solutions}
In this section we revisit the  example first discussed
in section 2.1.  We will show that the rest masses of
 particles may vary in a cosmological frame which employs a
comoving coordinate system, and make some comments about the observability
of the  geometric force.

Since the 4D metric of (\ref{404}) has a non-trivial
$\ell$-dependence, $\Lambda_{\mbox{{\scriptsize \em eff}}}$ is not
constant.  Furthermore, there is a non-trivial mass-variation, and equation
(\ref{c31}) reduces to the following rest-mass variation (by
identifying $\ell$ with $m$):
\be \label{c38}
\frac{\dot{m}}{m}=\pm \left(\frac{1-\al}{\al}\right) \frac{1}{t}.
\ee
Assuming $t \sim 10^{9}\, yr$ as an order of magnitude for the age of
the Universe \cite{Chaboyer1998a,Chaboyer1997a}, we find that for
$\alpha \lesssim 1$ the variation of rest masses is less than $10^{-11}
\,yr^{-1}$ which is consistent with the classical tests of 4D general
relativity \cite{Overduin1997a,Bekenstein1977a}.

The acceleration equation for the Ponce de Leon metric is simplified by
the comoving coordinate system.  In general, the assumption that the
spatial velocities are constant ($u^i=0$), implies that the scalar
field can only depend on time, so $\phi=\phi(t)$.  Thus we can
conclude that any 5D metric in the canonical form of (\ref{201}),
which has the 4D section $g_{\al\bet}(x^\Sigma, \ell)$ written in
comoving coordinates with a time-dependent scalar field, will not
impart a geometric force and the motion will be 4D geodesic.  This
applies to any spatially isotropic and homogeneous model (i.e., most
cosmological models) wherein comoving coordinates may be employed.

\subsection{Example B: Shell-like Solutions}
For the class of solutions (\ref{Shell}), $\phi=1$ and thus 
\be
\frac{\dot{\ell}}{\ell}=\pm\frac{\sqrt\Upsilon}{L} \quad \Leftrightarrow \qquad 
\dot \ell \propto e^{\pm \sqrt\Upsilon(s-s_0)/L}. 
\ee
Hence, this allows $\dot\ell\rightarrow 0$ at late times and so the
geometric force acting on the 4D particle motion can exponentially
decay in proper time, $s$; particles following this motion will asymptote
toward $\ell=\ell_0$.  

In order to explicitly calculate the geometric force (\ref{c34}), we
first need to determine the four--dimensional velocities $u_\alpha$.
These can be obtained either from solving (\ref{c33}) or deriving them
from the 5D geodesics.  Since the five-dimensional manifold is Riemann
flat, the 5D geodesics are easily obtainable, and it can be shown that
the 5D null geodesics are satisfied by
\beqn
\str{t} &=& \frac{L^2E}{\ell^2A^2}, \\
\str{\varphi} &=& \frac{L^2 {\cal J}}{\ell^2r^2C^2},\\
\str{r}&=& \frac{1}{lB^2}\left(L\sqrt{Q_1}-\varepsilon rB\sqrt{Q_2}\right),\\
\str{\ell}&=& \frac{1}{LB}\left(rB\sqrt{Q_1}+\varepsilon L \sqrt{Q_2}\right),
\eeqn
where
\beqn
Q_1 &=& \zeta_0^2 - \frac{L^2{\cal J}^2}{\ell^2r^2C^2},\\
Q_2 &=& \frac{L^2E^2}{\ell^2A^2}-\zeta_0^2,
\eeqn
$\varepsilon^2=1$, and $\{E,{\cal J},\zeta_0\}$ are integration constants in which $E$ may be
interpreted as the energy per unit rest mass and ${\cal J}$ is the
angular momentum per unit rest mass (note that we have consistently
chosen the declination angle to be $\theta=\pi/2$ with $\str{\theta}=0$).

To obtain the 4D velocities, it is easy to show from the line element of this space--time
\begin{displaymath}
d\hat{s}^2 = \frac{\ell^2}{L^2}ds^2 - d\ell^2
\end{displaymath}
that for 5D null geodesics 
\be
\frac{d\hat{s}}{ds}= \frac{\ell\sqrt\Upsilon}{\phi L}\left(\str{\ell}\right)^{-1}, \qquad
\Longrightarrow \qquad \dot{x}^\alpha = \frac{\ell\sqrt\Upsilon}{\phi L} 
	\str{x^\alpha}\left(\str{\ell}\right)^{-1},
\ee
and therefore, the 4D velocities for massive particles ($\Upsilon=1$) are
\beqn
\dot{t} &=& \frac{L^2B E}{\ell A^2\left[rB\sqrt{Q_1}
		+\varepsilon L\sqrt{Q_2}\right]}, \\
\dot{\varphi} &=& \frac{L^2 B{\cal J}}{\ell r^2C^2\left[rB\sqrt{Q_1}
		+\varepsilon L\sqrt{Q_2}  \right]},\\
\dot{r}&=& \frac{1}{B}\left(\frac{L\sqrt{Q_1}-\varepsilon rB\sqrt{Q_2}}{rB\sqrt{Q_1}+\varepsilon L\sqrt{Q_2}}\right).
\eeqn
It is apparent from these velocities that there will in general be a
geometric force acting on massive particles,
\be
\label{drag}
u^\bet \del_\bet\, u^\al = -2h^{\alpha \gamma}\left[\delta^0_\gamma ~\left|k_2\right|~A\dot{t}-\delta^3_\gamma ~|k_3|~\frac{L}{\ell}~rC\dot\varphi \right]\frac{\dot\ell}{\ell}.
\ee
As is evident in (\ref{drag}), we clearly see a drag force in the
$\phi$ direction, which is odd for spherically-symmetric solutions.
However, this is not unique to this particular solution and from
equation (\ref{c34}) it is apparent that there will in general be such
drag terms as long as the angular part of the metric has dependence in
the extra coordinate.

\setcounter{equation}{0}
\section{Final Comments}  
By retaining the extra coordinate $x^4=\ell$ in 5D Kaluza-Klein
gravity we have seen that a 5D vacuum induces non-trivial matter on 4D
hypersurfaces $\ell=\ell_o$, in which we retrieve a component which
acts as a cosmological ``constant'', a component which can be modeled
as a fluid and a scalar field contribution (if present).  Rather than
considering all three components as a single fluid source, we feel it
is important to keep the components distinct because of the close
connection between 5D vacuum relativity and 4D general relativity with
matter and a scalar field (see \cite{Billyard1997a}), and because the
discrepancies between gravitation and inertial masses do not arise
when one considers the scalar field separately (see
\cite{Sajko2000a}).  In particular, we use the eigenvalues and
eigenvectors of the induced energy--momentum tensor to properly
interpret the induced matter.  However, the induced stress--energy
tensor does not in general uniquely determine the matter content and
the interpretation chosen (for example, whether to model it as with
heat conduction or not, etc.) can lead to quite different kinematic
quantities.

We have shown that the assumption of 5D null geodesics can lead to a
variable rest mass for massive particles, once we identify the extra
dimension with mass.  The existence of a scalar field could be
inferred from particle motion in the coming Satellite Test of the
Equivalence Principle (STEP) \cite{Reinhard1993a}, and consequently
any such scalar field would place constraints on the rest mass
variation.  The acceleration for null particles remained the same as
in regular 4D relativity, but the motion for massive particles was
augmented by an additional force.  This force has a contribution from
a scalar field and crucially depends on the existence of the extra
dimension.  This motion was investigated for the Ponce de Leon class
of solutions with a particular extra coordinate dependence that
induced a time-varying cosmological constant $\Lambda\sim t^{-2}$.
For this metric there is no fifth force due to the nature of the
comoving coordinate system.  Indeed, for any metric which allows
comoving coordinates there will be no fifth force and so a majority of
simple cosmological models (e.g., non-tilting models, etc.) in general
will appear to allow motion which is geodesic in 4D.  For the
Shell--like solutions, the 4D motion of particles derived from 5D null
geodesics indeed asymptote to 4D geodesics since $\dot\ell \rightarrow
0$ exponentially.  In this example, the induced--matter is not what
would be observed by particles traveling along 5D null geodesics
until at late times.  Indeed, the two paradigms (induced matter and 5D
geodesics) are distinct.  One cannot say in general that particles
traveling along 5D null geodesics will observe the induced matter
derived from $\ell=\ell_0$ hypersurfaces, but as the Shell--like
solutions demonstrate, it may be possible that at some point in the
particle's path (early proper times, late proper times, etc) that the
two theories will indeed coincide.

Furthermore, any ``angular'' drag force terms which arise,
as demonstrated in the Shell--like solutions, would induce motion which
deviates from that of classical 4D motion and thus provides constraint
on this theory.  For example, the absence of drag terms in the angular
direction in 4D motions suggest that an appropriate 5D metric should
be independent of the extra coordinate in the angular components.  It
seems that we should turn to $\ell$-dependent analogues of the
Schwarzschild metric to observe and test any deviations from the
classical tests of GR due to the fifth force.  Work on this is
under way, and we expect to relate 5D dynamics to the upcoming Space
Test of the Equivalence Principle.  Finally, it is important to note
that the 4D velocities of the test particles derived from the 5D
motion {\em do not} correspond to the velocities of the induced
fluids, although this has often been assumed in the past (for a full
discussion, see \cite{Billyard1997b}), but rather they should be
interpreted as the velocity of a test particle traveling through the
fluid.  This is a consistent interpretation within regular GR in which
geodesics are assumed for test particles traveling through a fluid
\cite[Ch. 5.3]{Wald1984a}

\begin{center}{\bf Acknowledgments}\end{center}
The authors would like to thank G. S. Hall for useful
comments.  APB has been supported by the Natural Sciences and
Engineering Research Council of Canada and WNS by the Ontario 
Graduate Scholarship Program.

\appendix
\renewcommand{\theequation}{\mbox{\Alph{section}.\arabic{equation}}}
\renewcommand{\beqn}{\setcounter{saveeqn}{\value{equation}}%
	\stepcounter{saveeqn}\setcounter{equation}{0}%
	\renewcommand{\theequation}
	{\mbox{\Alph{section}.\arabic{saveeqn}\alph{equation}}}%
	\begin{eqnarray} }%

\renewcommand{\eeqn}{\end{eqnarray}\setcounter{equation}{\value{saveeqn}}%
	\renewcommand{\theequation}{\Alph{section}.\arabic{equation}}}%

\renewcommand{\mainlabel}[1]{\renewcommand{\theequation}%
	{\Alph{section}.\arabic{saveeqn}}\label{#1}%
	\renewcommand{\theequation}{\mbox{\Alph{section}.\arabic{saveeqn}%
	\alph{equation}}}}%

\setcounter{equation}{0}
\section{Extracting Kinematic Quantities from the Energy--Momentum Tensor\label{eigen}}
This appendix describes how kinematic variables can be obtained from
eigenvalues and eigenvectors of the energy-momentum tensor for
non-perfect fluids, generalizing the work found in \cite[Chapter
5.1]{Kramer1980a}.  Various energy-momentum tensors have physical
restrictions based on their \Seg type \cite{Hall1984a,Hall1993a}.  One
must insist that the metric's determinant be Lorentzian
(${\rm det}\,g_{\al\beta}<0$), which eliminates \Seg types $\{22\}$ and $\{4\}$
\cite{Hall1984a,Hall1993a}.  Furthermore, the strong energy condition,
$T^\al_\beta t^\beta t_\al>0$ (where $t^\beta$ is any time--like vector) eliminates \Seg
types $\{z,\bar z,1,1\}$ and $\{3,1\}$.  Finally, the {\em only} \Seg
type that admits a time--like eigenvector is $\{1,1,1,1\}$ and its
degeneracies, which is necessary for considering fluids with a
time--like velocity, $u^a$. Thus, the main focus here will be on
energy-momentum tensors of \Seg type \{1,1,1,1\} and its degeneracies.

We begin by assuming the standard non-perfect fluid energy momentum tensor, 
\be
\label{canon}
T^\al_\beta=\left(\mu+p\right)u^\al u_\beta -p\delta^\al_\beta 
	+u^\al q_\beta+u_\beta q^\al+\pi^\al_\beta,
\ee
where $u^\alpha$ is the fluids velocity field, $\mu$ is the fluid's
energy density, $p$ is the averaged pressure, $q^\alpha$ is the heat
conduction vector and $\pi^\alpha_\beta$ is the anisotropic pressure
tensor.  These quantities are constrained by
\be
u^\beta\pi^\al_\beta=0, \quad u^\beta q_\beta=0, \quad \pi^\al_\al=0, 
	\quad u^\al u_\al=1, \quad q^\al q_\al\equiv -q^2.
\ee
One could equally write
\be
T^\al_\beta=\left(\mu+p-\zeta\theta\right)u^\al u_\beta -(p-\zeta\theta) 
	\delta^\al_\beta +u^\al q_\beta+u_\beta q^\al-2\eta\sigma^\al_\beta,
\ee
to introduce the velocity's shear tensor, $\sigma^\al_\beta$, its
expansion scalar, $\theta$, as well as the fluid's bulk viscosity
coefficient, $\zeta$, and its shear viscosity, $\eta$.  One must be
careful here, since it is the velocity alone which determines
$\sigma^\al_\beta$ and $\theta$, and so if one may indeed have
$\sigma^\al_\beta \neq0$ even if it was initially assumed to be zero.
The expansion term may be ``absorbed'' by letting $p=\tilde p
+\zeta\theta$, and so this term can never be determined from the
eigenvalues of $T^\al_\beta$ alone.  Therefore, the form (\ref{canon})
will be used throughout.  Should $\sigma^\al_\beta\neq0$ and
$\pi^\al_\beta\propto\sigma^\al_\beta$, then the shear viscosity
coefficient, $\eta$ can also be calculated.

If one assumes that $\pi^\al_\beta\neq 0$, then it will have three
eigenvectors associated with its principle axes: \vup{\al} ($i=\{1,2,3\}$),
 where $\vup{\al}\vdown{\al}=-1$, and
$\vup[i]{\al}\vdown[j]{\al}=0$ for $i\neq j$.  Therefore, we write
$\pi^\al_\beta =\sum_i \pi_i\vup{\al}\vdown{\beta}$ (summation over for $i$
will remain explicit).  Since $\pi^\al_\al=0$ then $\pi_1+\pi_2+\pi_3=0$
and there are only two independent values for
$\{\pi_1,\pi_2,\pi_3\}$.  Hence, we will assume the the three
space--like eigenvectors can be written in terms of these three
vectors.  Should there be no anisotropies, \vup[1]{a} can be used to
denote the direction of $q^\al = q_1\vup[1]{\al}$ and \vup[2]{\al},
\vup[3]{\al} will be the (eigen)vectors perpendicular to $q^\al$ and
$u^\al$.

{\em All} eigenvectors, $y^\al$, will contain $p$, so to reduce
computation we will define
\beqn
\tilde T^\al_\beta& \equiv & T^\al_\beta + p\delta^\al_\beta,\\
\tilde\lambda&\equiv& \lambda + p,
\eeqn
where $\tilde \lambda$ is defined by 
\be
\label{TheEigen}
\tilde T^\al_\beta y^\beta=\tilde\lambda y^\al.
\ee

Hence, for $q^\al=\sum_i q_i\vup{\al}$ and $\pi^\al_\beta=\sum_i \pi_i 
\vup{\al}\vdown{\beta}$, we have:
\beqn
\label{ueigen}
\tilde T^\al_\beta u^\beta & = & (\mu+p) u^\al + q_1\vup[1]{\al}+ 
	q_2\vup[2]{\al}+ q_3\vup[3]{\al}, \\
\label{veigen}
\tilde T^\al_\beta \vup{\beta} & = & -\left[q_i u^\al+\pi_i\vup{\al}\right] \\
\label{qeigen}
\tilde T^\al_\beta q^\beta & = & -\left[q^2 u^\al + \sum_{i=1}^3 q_i \pi_i \vup{\al}\right],
\eeqn
where $q^2=q_1^2+q_2^2+q_3^2$.  If one multiplies each equation of
(\ref{veigen}) with $q_i$ and sum, one yields equation (\ref{qeigen})
and so the last may be omitted when considering $q^\al\neq0$,
$\pi^\al_\beta \neq0$.  However, in the event that $\pi^\al_\beta=0$
or $q^\beta\pi^\al_\beta=0$, then one may take \vup[1]{\al} as the
direction of $q^\al$ and the other two perpendicular to \vup[1]{\al},
and so the first equation of (\ref{veigen}) may be replaced by
(\ref{qeigen}).

In general, we seek eigenvectors of the form
\be
\xup[\ ]{\al} = a u^\al + b \vup[1]{\al} + c \vup[2]{\al} + d \vup[3]{\al} ,
\ee
where $\{a,b,c,d\}\in \mR$.  Hence, (\ref{TheEigen}) yields
the four equations:
\beqn
\label{aeqn}
a\tilde\lambda & = & a(\mu+p) -bq_1-cq_2-dq_3, \\
\label{beqn}
b\tilde\lambda & = & aq_1 -b\pi_1, \\
\label{ceqn}
c\tilde\lambda & = & aq_2 -c\pi_2,\\
\label{deqn}
d\tilde\lambda & = & aq_3 -d\pi_3.
\eeqn
Here, we have four equations for five unknowns
$\{\tilde\lambda,a,b,c,d\}$ and so we may arbitrarily set one to a
particular value (say, to normalize the vector).  This is a reflection
of the fact that eigenvectors can be arbitrarily scaled without
affecting (\ref{TheEigen}).  Although this may make the system
determined, we need to express the seven quantities
$\{\mu,p,q_1,q_2,q_3,\pi_1,\pi_2\}$ in terms of the four eigenvalues,
and so we would then need auxiliary equations (at most 3) to specify
all parameters.  However, we shall only consider here two cases,
$\pi^\al_\beta=0$, $q^\al\neq0$ and $\pi^\al_\beta\neq0$ and
$q^\al=0$, and for these cases the system is closed.

\subsection{Case 1: $\pi^a_b=0$}

For $\pi^\al_\beta=0$, it has been shown that $T^\al_\beta$ has to be of \Seg
type $\{1,1,(1,1)\}$ \cite{Hall1993a} (providing that
$(\mu+p)^2+4q^2>0$), with two degenerate eigenvalues.  Here, the
eigenvectors \vup[2,3]{\al} will be orthogonal to $u^\al$ and $q^\al$ with
eigenvalues $\lambda_2=\lambda_3=- p$.  We then need to find
the two other eigenvectors $\xup{\al}$ and their corresponding
eigenvectors $\tilde\lambda_\pm$.  In this case, we may let
$b\rightarrow bq_1$ ($q_1^2=q^2$) and consider only (\ref{aeqn}) and
(\ref{beqn}):
\begin{eqnarray*}
a\tilde\lambda_\pm & = & a(\mu+p) -b q^2 \\
b\tilde\lambda_\pm & = & a.
\end{eqnarray*}
The solutions to these equations are 
\beqn
\frac{a_\pm}{b_\pm} &=& \tilde\lambda_\pm = \half(\mu+p) 
	\pm\half\sqrt{(\mu+p)^2-4q^2} \\
\lambda_\pm & = & -\half(p-\mu) \pm\half \sqrt{(\mu+p)^2-4q^2}.
\eeqn
Defining,
\beqn
\Delta\lambda &\equiv& \lambda_+-\lambda_-=\sqrt{(\mu+p)^2-4q^2},\\
\bar\lambda &\equiv & \half\left(\lambda_++\lambda_-\right)=-\half(p-\mu),\\
\Lambda &\equiv &\lambda_2-\bar\lambda = -\half (p+\mu),
\eeqn
the magnitudes of \xup{a} are
\begin{eqnarray}
\frac{\xup[+]{2}}{b_+^2} &  & \left\{ {=\half\left[(\mu+p)^2
	-4q^2\right] 
	+\half (\mu+p)\sqrt{(\mu+p)^2-4q^2}}\atop{
	 =\half \Delta\lambda^2 - \Lambda\Delta\lambda \hfill} \right., \\
\nonumber && \\
\frac{\xup[-]{2}}{b_-^2} &  & \left\{ {=\half\left[(\mu+p)^2
	-4q^2\right] 
	-\half  (\mu+p)\sqrt{(\mu+p)^2-4q^2}}\atop{
	 =\half \Delta\lambda^2 + \Lambda\Delta\lambda \hfill} \right., \\
\nonumber && \\
\xup{\al}\xdown[\mp]{\al} & = & 0,
\end{eqnarray}
where it may be shown that $|\Lambda| > \half
|\Delta\lambda|$ for $q^2>0$, and so \xup[+]{\al} is time--like ($\xup[+]{2}>0$) and
\xup[-]{\al} is space--like ($\xup[-]{2}<0$).  Hence, by defining
\be
b^{-2}_\pm = -\Lambda\Delta\lambda\pm\half\Delta\lambda^2 > 0,
\ee 
we normalize these vectors to  $\xup{\al}\xdown{\al}=\pm 1$.

The kinematic variables are, in terms of the
eigenvalues/eigenvectors,
\beqn
\mainlabel{heateqns}
\mu & = & \lambda_++\lambda_--\lambda_2, \\
p & = & -\lambda_2, \\
q^2 & = &\left[\lambda_2-\lambda_+\right]\left[\lambda_2-\lambda_-\right], \\ 
\nonumber \\ 
u^\al & = & \frac{1}{\Delta\lambda}\left(\frac{\xup[+]{\al}}{b_+}
	-\frac{\xup[-]{\al}}{b_-}\right), \\
q^\al & = & \half\left(\frac{\xup[-]{\al}}{b_-}+\frac{\xup[+]{\al}}{b_+}\right)
	+ \frac{\Lambda}{\Delta\lambda}\left(\frac{\xup[+]{\al}}{b_+}
	-\frac{\xup[-]{\al}}{b_-}\right).
\eeqn

As evident from the magnitudes of \xup{a} we have the 
following cases:
\begin{enumerate}
\item $(\mu+p)^2>4q^2$:  \xup[-]{\al} is time--like and 
	\xup[+]{\al} is space--like.  \Seg type $\{1,1,(1,1)\}$; physically 
	relevant. 
\item $(\mu+p)^2=4q^2$:  $\lambda_+=\lambda_-$ and 
	\xup{\al} are null.  \Seg type 	$\{2,(1,1)\}$. 
\item $(\mu+p)^2<4q^2$:  $\lambda_\pm$ and \xup{\al} 
	are complex. \Seg type $\{z,\bar z,(1,1)\}$. 
\end{enumerate}

\subsection{Case 2: $q^\al=0$}

This case is fairly simple, since $\vup{\al}u_\al=0$ and
$\vup[i]{\al}\vdown[j]{\al}=0$, and so it is quite apparent that $\{u^\al,
\vup[1]{\al}, \vup[2]{\al}, \vup[3]{\al}\}$ are eigenvectors (see equations
(\ref{ueigen}) and (\ref{veigen})).  Denoting $\lambda_0$ to be the eigenvalue associated with $u^\al$, one has

\beqn
\mainlabel{noheateqns}
\mu & = & \lambda_0 \\
p   & = & -\third\left(\lambda_1+\lambda_2+\lambda_3\right) \\
\pi_i & = & -\lambda_i-p = -\left[
	\lambda_i -\third\sum_{j=1}^3 \lambda_j\right]
\eeqn
If one finds $\sigma^\al_\beta\neq0$ and
$\sigma^\al_\beta\propto\pi^\al_\beta$ then the shear viscosity
coefficient may be determined via
\be 
\eta=-\frac{\pi_1}{2\sigma_1}=-\frac{\pi_2}{2\sigma_2}
=-\frac{\pi_3}{2\sigma_3}, \ee
where $\sigma_i$ are the eigenvalues of the shear tensor.


\begin{thebibliography}{10}

\bibitem{Overduin1997a}
Overduin, J. and Wesson, P.~S. (1997).
\newblock {\em Phys. Rep.\/} 283, 303.

\bibitem{Sajko1998a}
Sajko, W.~N., Wesson, P.~S., and Liu, H. (1998).
\newblock {\em J. Math. Phys\/} 39, 2193.

\bibitem{Rippl1995a}
Rippl, S., Romero, C., and Tavakol, R. (1995).
\newblock {\em Class. Quantum Grav.\/} 12, 2411.

\bibitem{Romero1996a}
Romero, C., Tavakol, R., and Zalaletdinov, R. (1996).
\newblock {\em Gen. Rel. Grav.\/} 28, 365.

\bibitem{Billyard1996c}
Billyard, A.~P. and Wesson, P.~S. (1996).
\newblock {\em Gen. Rel. Grav.\/} 28, 129.

\bibitem{Billyard1996b}
Billyard, A.~P. and Wesson, P.~S. (1996).
\newblock {\em Phys. Rev. D\/} 53, 731.

\bibitem{Mashhoon1994a}
Mashhoon, B., Liu, H., and Wesson, P.~S. (1994).
\newblock {\em Phys. Lett. B\/} 331, 305.

\bibitem{Billyard1997a}
Billyard, A.~P. and Coley, A.~A. (1997).
\newblock {\em Mod. Phys. Lett. A\/} 12, 2121.

\bibitem{Wesson1996a}
Wesson, P.~S., {Ponce de Leon}, Liu, H., Masshoon, B., Kalligas, D., Everitt,
  C. W.~F., Billyard, A., Lim, P., and Overduin, J. (1996).
\newblock {\em Int. J. Mod. Phys. A\/} 11, 3247.

\bibitem{Wesson1999a}
Wesson, P.~S. (1999).
\newblock {\em Space, Time, Matter: Modern {K}aluza-{K}lein Theory\/}.
\newblock (World Scientific, River Edge, New Jersey).

\bibitem{Wesson1984a}
Wesson, P.~S. (1984).
\newblock {\em Gen. Rel. Grav.\/} 16, 193.

\bibitem{Mashhoon1998a}
Mashhoon, B., Wesson, P.~S., and Liu, H. (1998).
\newblock {\em Gen. Rel. Grav.\/} 30, 555.

\bibitem{Ponce1988a}
{Ponce de Leon}, J. (1988).
\newblock {\em Gen. Rel. Grav.\/} 20, 539.

\bibitem{Lopez1996a}
Lopez, J.~L. and Nanopoulos, D.~V. (1996).
\newblock {\em Mod. Phys. Lett. A\/} 11, 1.

\bibitem{Endo1977a}
Endo, M. and Fukui, T. (1977).
\newblock {\em Gen. Rel. Grav.\/} 8, 833.

\bibitem{Overduin1998a}
Overduin, J. and Cooperstock, F.~I. (1998).
\newblock {\em Phys. Rev. D\/} 58, 043506.

\bibitem{Wesson1992a}
Wesson, P.~S. (1992).
\newblock {\em Astrophys. J.\/} 394, 19.

\bibitem{Sajko1999b}
Sajko, W.~N. (1999).
\newblock {\em Phys. Rev. D\/} 60, 104038.

\bibitem{Wesson1998a}
Wesson, P.~S. and Liu, H. (1998).
\newblock {\em Phys. Lett. B\/} 432, 266.

\bibitem{Hall1984a}
Hall, G.~S. (1984).
\newblock {\em Arab. J. Sci. Eng.\/} 9, 87.

\bibitem{Wesson1997a}
Wesson, P.~S., Mashhoon, B., and Liu, H. (1997).
\newblock {\em Mod. Phys. Lett. A\/} 12, 2309.

\bibitem{Chaboyer1998a}
Chaboyer, B. (1998).
\newblock {\em Phys. Rept.\/} 307, 23.

\bibitem{Chaboyer1997a}
Chaboyer, B., Demarque, P., Kernan, P.~J., and Krauss, L.~M. (1998).
\newblock {\em Astrophys. J.\/} 494, 96.

\bibitem{Bekenstein1977a}
Bekenstein, J.~D. (1977).
\newblock {\em Phys. Rev. D\/} 15, 1458.

\bibitem{Sajko2000a}
Sajko, W.~N. (2000).
\newblock {\em Int. J. Mod. Phys. D\/} 9, 445.

\bibitem{Reinhard1993a}
Reinhard, R., Jafry, Y., and Laurance, R. (1993).
\newblock {\em Euro. Space Agency Jour.\/} 17, 251.

\bibitem{Billyard1997b}
Billyard, A.~P. and Coley, A.~A. (1997).
\newblock {\em Mod. Phys. Lett. A\/} 12, 2223.

\bibitem{Wald1984a}
Wald, R.~M. (1984).
\newblock {\em General Relativity\/}.
\newblock (The University of Chicago Press, Chicago, Illinois).

\bibitem{Kramer1980a}
Kramer, D., Stephani, H., Herlt, E., and MacCallum, M. A.~H. (1980).
\newblock {\em Exact Solutions of Einstein's Field Equations\/}.
\newblock (Cambridge University Press, Cambridge).

\bibitem{Hall1993a}
Hall, G.~S. (1993).
\newblock {\em Symmetries in General Relativity\/}.
\newblock Monograph series, Centro Brasileiro de Pesquisas F\'{i}icas.
\newblock CBPF-MO-001/93.

\end{thebibliography}

\end{document}